\documentclass{aa}
\usepackage[varg]{txfonts}
\usepackage[]{graphicx}
\usepackage{txfonts}

\begin{document}

\title{Photophoresis in the circumjovian disk and its impact on the orbital configuration of the Galilean satellites}
\titlerunning{Photophoresis in the Circumjovian Disk}

\author{
Sota Arakawa \inst{\ref{inst1}}
\and
Yuhito Shibaike \inst{\ref{inst2},\ref{inst3}}
}

\institute{
Department of Earth and Planetary Sciences, Tokyo Institute of Technology, Meguro, Tokyo, 152-8551, Japan\\
\email{arakawa.s.ac@m.titech.ac.jp} \label{inst1}
\and
Physikalisches Institut \& NCCR PlanetS, Universitaet Bern, CH-3012 Bern, Switzerland \label{inst2}
\and
Earth-Life Science Institute, Tokyo Institute of Technology, Meguro, Tokyo, 152-8550, Japan \label{inst3}
}

\date{Received / Accepted}

\abstract
{
Jupiter has four large regular satellites called the Galilean satellites: Io, Europa, Ganymede, and Callisto.
The inner three of the Galilean satellites orbit in a 4:2:1 mean motion resonance; therefore their orbital configuration may originate from the stopping of the migration of Io near the bump in the surface density distribution and following resonant trapping of Europa and Ganymede.
The formation mechanism of the bump near the orbit of the innermost satellite, Io, is not yet understood, however.
Here, we show that photophoresis in the circumjovian disk could be the cause of the bump, using analytic calculations of steady-state accretion disks.
We propose that photophoresis in the circumjovian disk could stop the inward migration of dust particles near the orbit of Io.
The resulting dust depleted inner region would have a higher ionization fraction, and thus admit increased magnetorotational-instability-driven accretion stress than the outer region.
The increase of the accretion stress at the photophoretic dust barrier would form a bump in the surface density distribution, halting the migration of Io.
}


\maketitle

\section{Introduction}

The gaseous planets in the solar system, Jupiter and Saturn, have large regular satellites whose radii are larger than 1000 km, and their orbits are on the equatorial plane of the central planets.
It is widely accepted that the gaseous planets were surrounded by a circumplanetary disk and large satellites formed in the disks \citep[e.g.,][]{Lunine+1982,Mosqueira+2003,Canup+2006}.
Several hydrodynamic simulations revealed that circumplanetary disks can be formed as byproducts of the formation and growth of gaseous planets via mass accretion from the parent protoplanetary disk \citep[e.g.,][]{Machida+2010,Tanigawa+2012}.

\citet{Canup+2002} proposed the gas-starved disk model whose disk mass is regulated by the steady-state viscous accretion onto the central planet.
\citet{Canup+2006} then performed $N$-body simulations of satellite formation in the gas-starved disk and revealed that the total mass of the satellites in circumplanetary disks is successfully reproduced.
Moreover, \citet{Sasaki+2010} found that the number, masses, and orbits of the Galilean satellites can be reproduced when the disk had an inner cavity at their formation era.

Jupiter has four large regular satellites called the Galilean satellites: Io, Europa, Ganymede, and Callisto.
Saturn, in contrast, has only one large regular satellite, Titan, and many 100--1000 km-sized satellites.
The inner three Galilean satellites orbit in a 4:2:1 mean motion resonance.
One of the mechanisms for reproducing this orbital configuration is the stopping of the migration of Io near the bump in the surface density distribution and the subsequent resonant trapping of Europa and Ganymede \citep[e.g.,][]{Sasaki+2010,Fujii+2017}.

There have been two hypotheses to explain the formation of the bump in the surface density distribution of circumplanetary disks.
\citet{Fujii+2017} revealed that in the early stage of disk evolution, the disk temperature is high and the sublimation of silicate dust particles occurs within a certain orbital radius.
This absence of dust particles changes the opacity \citep[e.g.,][]{Bell+1994}, causing the formation of the bump in the surface density distribution of the circumjovian disk.
One problem in this scenario is whether the satellites can survive against the radial drift in the late stage of the circumjovian disk when the temperature of the disk edge is sufficiently lower than the sublimation temperature.
\citet{Fujii+2017} noted that the fate of satellites depends on the time evolution of the mass accretion rate onto Jupiter after the formation of planetary gas in the solar nebula \citep[e.g.,][]{Durmann+2015,Kanagawa+2018}.
Another problem is that icy satellites, at a minimum, cannot survive longer than the disk dissipation timescale in such a high-temperature disk, because the temperature at the current location of icy satellites is about 1000 K at this stage (see Appendix \ref{AppA}).
We found that icy satellites at the ambient temperature of 500 K might be evaporated within 0.1 Myr, even if the satellite is as large as Ganymede.

On the other hand, \citet{Sasaki+2010} proposed that the inner cavity of the circumjovian disk could be formed by the magnetic coupling between the disk and young Jupiter.
This magnetic coupling can also explain why the spin rate of Jupiter is sufficiently slower than the break-up spin rate \citep[e.g.,][]{Takata+1996,Batygin2018}.
Nevertheless, one problem exists for this scenario: the spin period of young Jupiter should be longer than 40 h at the end of Galilean satellite formation, as the innermost satellite should have formed at the location of the inner cavity \citep[][Shibaike et al.\ submitted]{Ogihara+2012}.
This spin period is four times longer than the current spin period (approximately 10 h); therefore, the planetary radius of young Jupiter may be at least twice as large as the current planetary radius (see Appendix \ref{AppB}), although several evolutionary calculations have indicated that the planetary radius of young Jupiter might be approximately 1.5 times the current radius at 1 Myr after the formation of Jupiter \citep[e.g.,][]{Fortney+2011}.

In this study, we propose another mechanism to create a bump in the surface density distribution of the circumjovian disk---the depletion of fine dust particles by photophoresis.
Photophoresis is the momentum transfer mechanism by gas molecules colliding with photo-irradiatied dust particles that exhibit a temperature gradient \citep[e.g.,][]{Krauss+2005,Loesche+2012}.
The temperature gradient within dust particles is thought to form via direct irradiation from the central planet because the inner region of the circumjovian disk could be sufficiently depleted of dust particles.
Thus, the inner region is optically thin.
The resulting dust depleted inner region would have a higher ionization fraction, and thus admit increased accretion stress driven by magnetorotational instability (MRI) \citep[e.g.,][]{Balbus+1991}.
This creates a bump in the gas surface density distribution and a pressure maximum at the boundary of the regions.
We show that the photophoretic force in the circumjovian disk would overcome the gas drag, and the inward migration of dust particles would be stopped.
We also reveal that this could form a pressure maximum and the bump in the surface density distribution of the circumjovian disk.
In our model, the location of the bump is near the orbital radius of Io, and we do not need to consider the significant radius anomaly of young Jupiter.
In addition, the temperature of the disk at the current location of icy satellites could be lower than the sublimation temperature of ${\rm H}_{2}{\rm O}$ ice, and the resonant-trapped icy satellites may survive in the disk.
Consequently, photophoresis in the circumjovian disk has the potential to explain the formation of the Galilean satellites.

\section{Models}

\subsection{Photophoresis and gas drag}

In the circumplanetary disk with gas and dust, whether or not dust particles drift inward is determined by the balance of the residual gravity caused by the gas drag, $F_{\rm D}$, and the photophoretic force, $F_{\rm Ph}$.
In gaseous disks, other light-induced forces, such as the radiation pressure and the Poynting--Robertson effect, are negligible for $\mu$m- or mm-sized dust particles \citep[][see also Appendix \ref{AppC}]{Wurm+2006}.

For perfectly absorbing spherical particles whose radius is smaller than the mean free path of the gas molecules $l_{\rm m}$, the photophoretic force $F_{\rm Ph}$ is given by \citep{Beresnev+1993,Wurm+2006}
\begin{equation}
F_{\rm Ph} = \frac{0.5 \pi a^{2} I p}{3 k_{\rm th} T / a + 12 \sigma_{\rm SB} T^{4} + p \sqrt{18 k_{\rm B} T / {( \pi m_{\rm g} )}}},
\label{eq1}
\end{equation}
where $a$ is the radius of dust particles, $I$ is the intensity of the radiative flux from the planet, $p$ is the gas pressure, $k_{\rm th}$ is the thermal conductivity of dust particles, $T$ is the temperature of the gas, $\sigma_{\rm SB} = 5.67 \times 10^{-5}\ {\rm erg}\ {\rm cm}^{-2}\ {\rm K}^{-4}\ {\rm s}^{-1}$ is the Stefan-Boltzmann constant, $k_{\rm B} = 1.38 \times 10^{-16}\ {\rm erg}\ {\rm K}^{-1}$ is the Boltzmann constant, and $m_{\rm g} = 3.9 \times 10^{-24}\ {\rm g}$ is the mean mass of gas molecules.
We set $a = 3\ {\mu}{\rm m}$, which is consistent with the result of recent in-situ measurements of dust particles of the comet 67P/Churyumov-Gerasimenko \citep[e.g.,][]{Bentley+2016} and the estimation of the radius of constituent particles called monomers in the protoplanetary disk around the young protostar HL Tau \citep[e.g.,][]{Okuzumi+2016}.
The thermal conductivity of compact dust particles with no voids is about a few $10^{-1}\ {\rm W}\ {\rm m}^{-1}\ {\rm K}^{-1}$ for organic polymers \citep[e.g.,][]{Choy1977} and $\sim 1\ {\rm W}\ {\rm m}^{-1}\ {\rm K}^{-1}$ for glass beads \citep[e.g.,][]{Sakatani+2017}.
Then we set $k_{\rm th} = 3 \times 10^{4}\ {\rm erg}\ {\rm s}^{-1}\ {\rm cm}^{-1}\ {\rm K}^{-1}\ {( = 0.3\ {\rm W}\ {\rm m}^{-1}\ {\rm K}^{-1} )}$.

We checked the validity of the assumption that dust particles can be treated as perfectly absorbing particles for the radiation from the planet.
The effective temperature of young Jupiter is approximately 1000 K \citep[e.g.,][]{Fortney+2011}, thus the characteristic wavelength of the radiation from the young Jupiter is $\lambda \simeq 3\ {\mu}{\rm m}$.
Therefore, we can apply Equation (\ref{eq1}) for $\mu$m-sized dust particles whose size parameter $x \equiv 2 \pi a / \lambda$ is larger than unity, i.e., $a \gtrsim 0.5\ {\mu}{\rm m}$.
We also confirmed the validity of the assumption of $a < l_{\rm m}$ (see Appendix \ref{AppD}).

The rotation of dust particles is not considered in this study, although fast rotation results in a decrease in the photophoretic force.
For $\mu$m-sized dust particles, the main source of random rotation is Brownian motion.
We confirmed that the thermal rotation timescale is several orders of magnitude longer than the heat conduction timescale and the effect of rotation is negligible (see Appendix \ref{AppE}).

The gas in a circumplanetary disk usually rotates slower than the Keplerian velocity owing to the negative pressure gradient in the accretion disk.
In contrast, dust particles in the gas disk are not supported by the pressure gradient, and their stable orbital velocity is Keplerian; however, fine dust particles do rotate with the gas.
Therefore, there is a gas drag on dust particles, and the residual gravity $F_{\rm D}$ is given by \citep{Weidenschilling1977,Wurm+2006}
\begin{equation}
F_{\rm D} = - \frac{4 \pi a^{3}}{3} \frac{\rho_{\rm p}}{\rho_{\rm g}} \frac{{\rm d}p}{{\rm d}r},
\end{equation}
where $\rho_{\rm p} = 2\ {\rm g}\ {\rm cm}^{-2}$ is the density of dust particles, $\rho_{\rm g}$ is the gas density, and ${{\rm d}p}/{{\rm d}r}$ is the pressure gradient.

\begin{figure}
\centering
\includegraphics[width = \columnwidth]{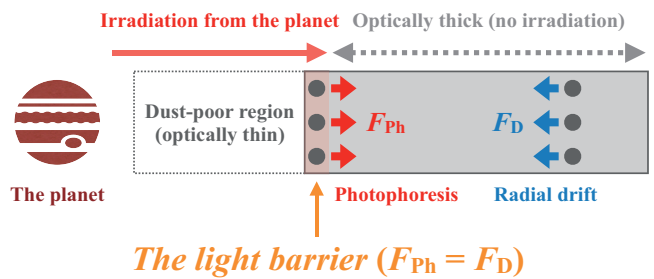}
\caption{
Schematic of the circumplanetary disk with a dust-poor inner region formed by photophoresis.
Both the pressure maximum and the bump in the surface density distribution are at the light barrier.
}
\label{fig1}
\end{figure}

In the outer region of the circumplanetary disk, the residual gravity $F_{\rm D}$ is larger than the photophoretic force $F_{\rm Ph}$, and the inward drift dominates, even in the presence of direct irradiation.
However, there is a turnover distance where $F_{\rm D} = F_{\rm Ph}$, and the inward drift is prevented.
Therefore, the inner region of the circumplanetary disk would be depleted of dust particles.
\citet{Wurm+2006} named this transition location {\it the light barrier} (see Figure \ref{fig1}).

\subsection{Circumplanetary disk model}

We model the structure of the circumplanetary disk as \citet{Fujii+2014} and \citet{Shibaike+2017}.
In this study, we assume the circumplanetary disk is a steady-state viscous accretion disk with a supply from the protoplanetary disk.
The gas surface density of the circumplanetary disk $\Sigma$ is given by
\begin{equation}
\Sigma = \frac{\dot{M}}{2 \pi r_{\rm b}} \frac{r^{3/2}}{\nu} {\left( - \frac{2}{9} r^{-1/2} + \frac{2}{3} r_{\rm b} r^{-3/2} \right)},
\end{equation}
where $\dot{M}$ is the mass accretion rate, $r_{\rm b}$ is the radius of the region where the gas falls in, $r$ is the distance from the planet, and $\nu$ is the kinematic viscosity.
The numerical simulation by \citet{Tanigawa+2012} found that $r_{\rm b} \simeq 20 R_{\rm J}$ when the planet has a mass of $0.4 M_{\rm J}$, where $R_{\rm J} = 7.15 \times 10^{4}\ {\rm km}$ and $M_{\rm J} = 1.9 \times 10^{30}\ {\rm g}$ are the planetary radius and mass of Jupiter \citep[see also][]{Fujii+2014}, respectively.
Assuming that $r_{\rm b}$ scales with the Hill radius of the planet, \citet{Shibaike+2017} suggested that $r_{\rm b} = 27 R_{\rm J}$ when the planet has a mass of $1 M_{\rm J}$.

The gas temperature at the disk midplane $T$ is given by \citep{Nakamoto+1994}
\begin{equation}
T = {\left[ \frac{9 \nu \Sigma {\Omega_{\rm K}}^{2}}{8 \sigma_{\rm SB}} {\left( \frac{3 \tau_{\rm R}}{8} + \frac{1}{2 \tau_{\rm P}} \right)} \right]}^{1/4},
\end{equation}
where $\Omega_{\rm K}$ is the Keplerian angular velocity, and $\tau_{\rm R}$ and $\tau_{\rm P}$ are the Rosseland and Planck mean optical depths, respectively.
The Keplerian angular velocity is given by $\Omega_{\rm K} = \sqrt{{G M} / {r^{3}}}$, where $G = 6.67 \times 10^{-8}\ {\rm dyn}\ {\rm cm}^{2}\ {\rm g}^{-2}$ is the gravitational constant, and $M$ is the mass of the central planet.
We set $M = M_{\rm J}$ in this study.
Then, the sound velocity $c_{\rm s}$ is $c_{\rm s} = \sqrt{k_{\rm B} T / m_{\rm g}}$, the gas scale height of the disk $h_{\rm g}$ is $h_{\rm g} = c_{\rm s} / \Omega_{\rm K}$, the gas density at the disk midplane is $\rho_{\rm g} = \Sigma / {( \sqrt{2 \pi} h_{\rm g} )}$, and the pressure at the disk midplane is $p = {\rho_{\rm g} k_{\rm B} T} / m_{\rm g}$.
We employ the standard $\alpha$-prescription for the kinematic viscosity $\nu$ as $\nu = \alpha {c_{\rm s}}^{2} / \Omega_{\rm K}$ \citep{Shakura+1973}.

The Rosseland mean optical depth $\tau_{\rm R}$ is given by
\begin{equation}
\tau_{\rm R} = \frac{\Sigma}{2} {\left( f_{\rm d} \kappa_{\rm R, d} + \kappa_{\rm R, g} \right)},
\end{equation}
where $f_{\rm d}$ is the scaling factor of the dust abundance, and $\kappa_{\rm R, d}$ and $\kappa_{\rm R, g}$ are the contributions of dust and gas to the Rosseland mean opacity of a disk with solar metallicity.
Similarly, the Planck mean optical depth $\tau_{\rm P}$ is given by
\begin{equation}
\tau_{\rm P} = \frac{\Sigma}{2} {\left( f_{\rm d} \kappa_{\rm P, d} + \kappa_{\rm P, g} \right)},
\end{equation}
where $\kappa_{\rm P, d}$ and $\kappa_{\rm P, g}$ are the contributions of dust and gas to the Planck mean opacity of a disk with solar metallicity.
The dust and gas opacities are taken from \citet{Henning+1996} and \citet{Freedman+2014}, respectively (see Appendix \ref{AppF}).

Because the dust particles in the solar nebula tend to settle down toward the midplane, the dust-to-gas mass ratio of the inflow gas is thought to be lower than the dust-to-gas mass ratio of the solar nebula \citep[e.g.,][]{Tanigawa+2012}.
In addition, the dust-to-gas mass ratio in the inner region and the outer region must differ by orders of magnitude by the light barrier.
Thus, we assumed that $f_{\rm d} = 10^{-1}$ for the outer region, and $f_{\rm d} = 10^{-6}$ for the inner region.

In this study, we assume that the abundance of nm- or sub-$\mu$m-sized dust particles is negligibly small, and $\kappa_{\rm R, d}$ and $\kappa_{\rm P, d}$ are determined by the abundance of $\mu$m-sized dust particles which affect the photophoretic force.
Thus we expect the drastic change of $f_{\rm d}$ at the light barrier.
We also neglect the turbulent diffusion of dust particles and consider the boundary at the light barrier to be sharp for simplicity.
We note that the diffusion of dust particles may be important if the diffused region becomes optically thick and the intensity of the radiative flux at the light barrier decreases.
Hence, future studies on this point are essential, although a detailed discusstion is beyond the scope of this study.

The depletion of dust also affects the turbulent viscosity of the disk.
\citet{Fujii+2011,Fujii+2014} found that the midplane of circumplanetary disks is in general MRI-inactive because the diffusion timescale of the magnetic field is short.
In the MRI-inactive region, it is likely that hydrodynamic instabilities, e.g., the convective overstability \citep[e.g.,][]{Klahr+2014} and subcritical baroclinic instability \citep[e.g.,][]{Lyra+2011}, could dominate the angular momentum transport with values of $\alpha \sim 10^{-3}$ \citep[e.g.,][]{Lyra2014}
\footnote{
We acknowledge, however, that these studies \citep{Klahr+2014,Lyra+2011} are for protoplanetary disks and we need to confirm whether or not hydrodynamic instabilities could dominate the angular momentum transport in the MRI-inactive region in future.}
.
In contrast, the disk might be (partly) MRI-active if it is significantly depleted of dust grains \citep[e.g.,][]{Turner+2014,Fujii+2014}, because the lower dust abundance leads to the higher ionization fraction of the disk gas.
This is true when the dominant fraction of the total surface area of dust particles (where free charges can stick) is in $\mu$m-sized dust particles which will be stopped at the light barrier.
In this case, the effective turbulent viscosity could be larger than the viscosity of the completely MRI-inactive disks.
Thus, we assumed $\alpha = 10^{-3}$ for the outer region and $\alpha = 10^{-2}$ for the inner region.
We note that the choice of absolute values of $\alpha$ does not change the results drastically; the increase in $\alpha$ at the light barrier is the key to generating the bump in the surface density distribution (see Appendix \ref{AppG}).

\section{Results}

\subsection{Location of the light barrier}

First, we calculate the location of the light barrier in the steady-state accretion disk.
At the location of the light barrier, the intensity of the radiative flux from the planet $I$ should be $I = L / {( 4 \pi r^{2} )}$, where $L$ is the luminosity of the planet.
As the rapid gas accretion was turned off and the planetary mass reached approximately $1 M_{\rm J}$, the luminosity of young Jupiter rapidly collapsed to $10^{-5} L_{\sun}$ \citep[e.g.,][]{Marley+2007}, where $L_{\sun} = 3.83 \times 10^{33}\ {\rm erg}\ {\rm s}^{-1}$ is the solar luminosity.
Then, the luminosity of young Jupiter was on the order of $10^{-5} L_{\sun}$ to $10^{-6} L_{\sun}$ within the first 10 Myr \citep[e.g.,][]{Fortney+2011}, and we set $L = 4 \times 10^{-6} L_{\sun}$ in our calculation.

We calculated the force ratio of the photophoretic force to the gas drag, $F_{\rm Ph} / F_{\rm D}$, by using both the normal model ($f_{\rm d} = 10^{-1}$ and $\alpha = 10^{-3}$) and the dust-poor model ($f_{\rm d} = 10^{-6}$ and $\alpha = 10^{-2}$).
The force ratio calculated from the normal model is ${\left( F_{\rm Ph} / F_{\rm D} \right)}_{\rm n}$, and the force ratio calculated from the dust-poor model is ${\left( F_{\rm Ph} / F_{\rm D} \right)}_{\rm p}$.
Then, we defined the location of the light barrier $r_{\rm lb}$ as the location where $\min {\left\{ {\left( F_{\rm Ph} / F_{\rm D} \right)}_{\rm n}, {\left( F_{\rm Ph} / F_{\rm D} \right)}_{\rm p} \right\}} = 1$, and we applied the normal model for the outer region whose distance is larger than $r_{\rm lb}$ and applied the dust-poor model for the inner region whose distance is smaller than $r_{\rm lb}$.

\begin{figure}
\centering
\includegraphics[width = \columnwidth]{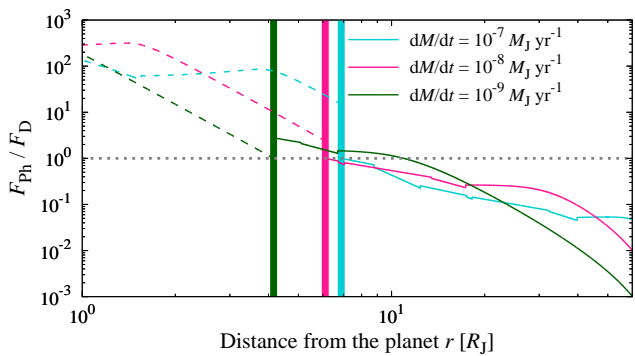}
\caption{
The force ratio $F_{\rm Ph} / F_{\rm D}$ under the assumption that $I = L / {( 4 \pi r^{2} )}$.
The solid lines correspond to the force ratio calculated from the normal model ($f_{\rm d} = 10^{-1}$ and $\alpha = 10^{-3}$), and the dashed lines are the dust-poor model ($f_{\rm d} = 10^{-6}$ and $\alpha = 10^{-2}$).
The vertical lines represent the location of the light barrier.
The line colors represent the mass accretion rate ($\dot{M} = 10^{-7} M_{\rm J}\ {\rm yr}^{-1}$ for blue, $\dot{M} = 10^{-8} M_{\rm J}\ {\rm yr}^{-1}$ for pink, and $\dot{M} = 10^{-9} M_{\rm J}\ {\rm yr}^{-1}$ for green).
As a reference, the current orbit of Io is $r = 5.9 R_{\rm J}$.
}
\label{fig2}
\end{figure}

Figure \ref{fig2} shows the force ratio $F_{\rm Ph} / F_{\rm D}$ under the assumption that the intensity of the radiative flux is $I = L / {( 4 \pi r^{2} )}$.
This estimate of $I$ is true at the light barrier.
We found that $r_{\rm lb} = 6.9 R_{\rm J}$, $6.1 R_{\rm J}$, and $4.2 R_{\rm J}$ for the cases of $\dot{M} = 10^{-7} M_{\rm J}\ {\rm yr}^{-1}$, $10^{-8} M_{\rm J}\ {\rm yr}^{-1}$, and $10^{-9} M_{\rm J}\ {\rm yr}^{-1}$, respectively.
Therefore, the photophoretic force stops the inward migration of dust particles, and the inner region of the circumjovian disk is depleted of fine dust particles.
Moreover, the location of this light barrier is close to the current orbit of the innermost satellite, Io ($r = 5.9 R_{\rm J}$).

We note that, for the case of $\dot{M} = 10^{-9} M_{\rm J}\ {\rm yr}^{-1}$, the force ratio is $F_{\rm Ph} / F_{\rm D} = 1$ not only at the light barrier ($r = 4.2 R_{\rm J}$) but also at $r = 10.9 R_{\rm J}$.
We speculate that the {\it true} location of the light barrier is between $4.2 R_{\rm J}$ and $10.9 R_{\rm J}$, although a detailed calculation of the dust and gas disk structure around the light barrier is needed to determine the location of the light barrier.

Behind the light barrier, the intensity of the radiative flux would decrease exponentially due to the increase in the optical depth from the light barrier.
Then, the photophoretic force might only work around the light barrier, and the structure of the outer region might not be affected.
However, the structure of the inner region must be affected by photophoresis, and our results hardly change.
We also note that $r_{\rm lb}$ hardly depends on the radius of dust particles when two conditions, $x \gg 1$ and $a \ll l_{\rm m}$, are satisfied (see Appendix \ref{AppH}).

\subsection{Pressure and surface density distributions}

\begin{figure}
\centering
\includegraphics[width = \columnwidth]{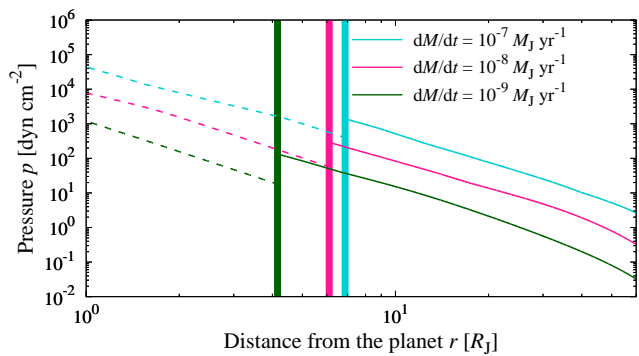}
\caption{
The pressure distribution of the disk.
The solid lines correspond to the pressure $p$ calculated from the normal model, and the dashed lines are the dust-poor model.
The vertical lines represent the location of the light barrier, and a pressure maximum is formed at the light barrier.
}
\label{fig3}
\end{figure}

Figure \ref{fig3} shows the pressure distribution of the circumjovian disk.
We found that a pressure maximum was formed at the light barrier.
This structure is originated from the change in the value of $\alpha$ at the light barrier in the steady-state accretion disk.
We speculate that the radial drift of pebbles and satellitesimals might stop at the pressure maximum \citep[e.g.,][]{Haghighipour+2003} and help the inner region to maintain its dust-poor environment.

\begin{figure}
\centering
\includegraphics[width = \columnwidth]{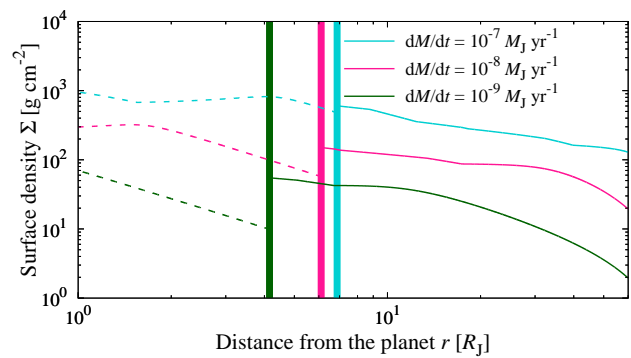}
\caption{
The surface density distribution of the disk.
The solid lines correspond to the surface density $\Sigma$ calculated from the normal model, and the dashed lines are the dust-poor model.
The vertical lines represent the location of the light barrier, and a bump in the surface density distribution is formed at the light barrier.
}
\label{fig4}
\end{figure}

Figure \ref{fig4} shows the surface density distribution.
We found that the bump in the surface density distribution is formed at the light barrier when the mass accretion rate is lower than $10^{-7} M_{\rm J}\ {\rm yr}^{-1}$.
The migration direction and speed depend on the mass of the satellite and the disk structure \citep[e.g.,][]{Tanaka+2002,Paardekooper+2011,Romanova+2019}.
\citet{Fujii+2017} found that the innermost satellite is stopped by the convergent migration to the bump.
From our calculation, a steep bump in the surface density distribution is formed at the light barrier.
Thus, the innermost satellite could be trapped around the light barrier.

Moreover, we confirmed the possibility of the formation of the resonant-trapped Galilean satellites by the following migration of Europa and Ganymede (see Appendix \ref{AppI}).
We found that the migration timescale is sufficiently longer than the critical timescale for resonance capture, and the adjacent satellites (i.e., Io--Europa and Europa--Ganymede) can be captured in the 2:1 mean motion resonance in the circumjovian accretion disk.

\subsection{Disk temperature}

\begin{figure}
\centering
\includegraphics[width = \columnwidth]{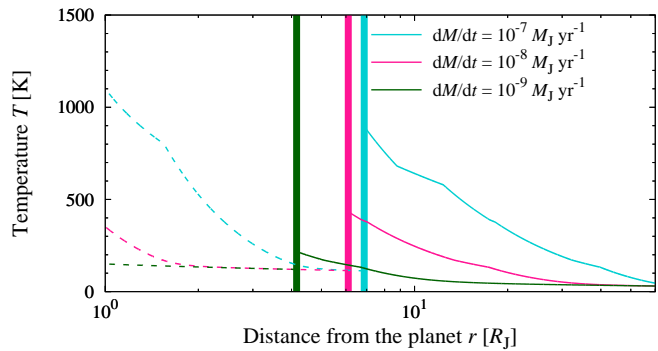}
\caption{
The temperature distribution of the disk.
The solid lines correspond to the temperature $T$ calculated from the normal model, and the dashed lines are the dust-poor model.
}
\label{fig5}
\end{figure}

Figure \ref{fig5} shows the temperature distribution.
We found that the temperature at the light barrier is lower than the sublimation temperature of silicate dust \citep[$\gtrsim 1500\ {\rm K}$;][]{Lodders2003} when $\dot{M} \lesssim 10^{-7} M_{\rm J}\ {\rm yr}^{-1}$.
Therefore, photophoresis works before the sublimation of fine dust particles.
Moreover, the temperature at the current location of Ganymede ($r = 15 R_{\rm J}$) becomes lower than the sublimation temperature of ${\rm H}_{2}{\rm O}$ ice ($\simeq 200\ {\rm K}$) when $\dot{M} \lesssim 10^{-8} M_{\rm J}\ {\rm yr}^{-1}$.
This result indicates that the Galilean satellites formed in the circumjovian disk with a mass accretion rate of $\lesssim 10^{-8} M_{\rm J}\ {\rm yr}^{-1}$.

We note, however, that the disk temperature depends on the assumed opacity \citep[e.g.,][]{Canup+2002}, and the opacity should be calculated from the size distribution and the true abundance of the dust particles.
The disk temperature is also dependent on the vertical structure of the accretion disk \citep[e.g.,][]{Mori+2019}, and the vertical distribution of dust particles might be altered by photophoresis itself \citep[e.g.,][]{McNally+2015,McNally+2017}.
Thus, future studies in these areas are essential.

\section{Conclusions}

The inner three Galilean satellites are in a 4:2:1 mean motion resonance, and their orbital configuration may be originated from the stopping of the migration of Io around the bump in surface density distribution, although the formation mechanism of the bump is under debate.
We found that photophoresis in the circumjovian disk stopped the inward migration of fine dust particles.
The resulting dust depleted inner region would have a higher ionization degree, and thus admit increased turbulent viscosity.
This also forms a bump in the surface density distribution, and the migration of the innermost large satellite terminated around the bump.
The location of the bump is near the current orbit of Io.
Hence, photophoresis is a strong candidate for the origin of the resonant-trapped Galilean satellites.

\begin{acknowledgements}
We thank Taishi Nakamoto for meaningful comments.
S.A.\ is supported by the Grant-in-Aid for JSPS Research Fellow (JP17J06861).
Y.S.\ is supported by a JSPS KAKENHI Grant (JP15H02065).
This work has been carried out within the frame of the National Centre for Competence in Research PlanetS supported by the Swiss National Science Foundation (SNSF).
The authors acknowledge the financial support of the SNSF.
\end{acknowledgements}

\bibliographystyle{aa}

\bibliography{references}

\begin{appendix}

\section{Evaporation of icy satellites}
\label{AppA}

If the ambient temperature is much higher than the sublimation temperature of ${\rm H}_{2}{\rm O}$ ice, then the icy satellites might be evaporated within the dissipation timescale of the circumplanetary disk.
Here, we estimate the evaporation timescale.
The mass and radius of the largest satellite, Ganymede, are $M_{\rm Ga} = 1.48 \times 10^{26}\ {\rm g}$ and $R_{\rm Ga} = 2.63 \times 10^{8}\ {\rm cm}$, and the evaporation timescale $\tau_{\rm evap}$ is evaluated as
\begin{equation}
\begin{split}
\tau_{\rm evap} &\simeq \frac{L_{\rm evap} M_{\rm sat}}{4 \pi {R_{\rm sat}}^{2} \sigma_{\rm SB} T^{4}} \\
                &\simeq 4.1 \times 10^{4} {\left( \frac{M_{\rm sat}}{M_{\rm Ga}} \right)} {\left( \frac{R_{\rm sat}}{R_{\rm Ga}} \right)}^{-2} {\left( \frac{T}{500\ {\rm K}} \right)}^{-4}\ {\rm yr},
\end{split}
\end{equation}
where $L_{\rm evap} = 2.7 \times 10^{10}\ {\rm erg}\ {\rm g}^{-1}$ is the latent heat of sublimation of ${\rm H}_{2}{\rm O}$ ice \citep{Tanaka+2013}, $M_{\rm sat}$ is the mass of the satellite, and $R_{\rm sat}$ is the radius of the satellite.
Therefore, the evaporation timescale is shorter than the dissipation timescale of the circumplanetary disk (a few Myr).
Moreover, the evaporation timescale is proportional to the radius $R_{\rm sat}$ and inversely proportional to the fourth power of the ambient temperature $T$, such that not only large satellites but also icy pebbles and satellitesimals cannot survive in the high-temperature environment.

\section{Inner cavity and corotation radius}
\label{AppB}

If the inner cavity is located around the orbit of Io ($r = 5.9\ R_{\rm J}$), the orbital configuration of the Galilean satellites can be reproduced \citep[][Shibaike et al.\ submitted]{Ogihara+2012}.
If the planetary magnetic field can be coupled effectively to the inner region of a circumplanetary accretion disk, then the disk could be disrupted at the radius $r_{\rm in}$ \citep[e.g.,][]{Koenigl1991}:
\begin{equation}
\begin{split}
r_{\rm in} &= {\left( \frac{\mu^{4}}{2 G M {\dot{M}}^{2}} \right)}^{1/7} \\
           &\simeq 5.2 R_{\rm J} {\left( \frac{B}{150\ {\rm G}} \right)}^{4/7} {\left( \frac{\dot{M}}{10^{-8} M_{\rm J}\ {\rm yr}^{-1}} \right)}^{- 2/7},
\end{split}
\end{equation}
where $\mu \sim B R^{3}$ is the dipole moment of the planet, $B$ is the surface magnetic field of the planet, and $R = R_{\rm J}$ is the planetary radius.
Although the current magnetic field of Jupiter is 7.8 G, according to the observations from the Juno spacecraft \citep{Bolton+2017,Connerney+2017}, young Jupiter might have a stronger magnetic field, on the order of 100 G \citep[e.g.,][]{Sanchez-Lavega2004,Reiners+2010}.
Thus, we can expect that $r_{\rm in} \simeq 5.9 R_{\rm J}$ could be achieved if the ionization degree of the circumjovian disk is high enough, and the planetary magnetic field is coupled to the disk.

However, there is another requirement to make an inner cavity in a steady-state disk.
To maintain its steady-state accretion, the corotation radius $r_{\rm co}$ should not be smaller than the radius of the inner cavity $r_{\rm in}$ \citep[e.g.,][]{Koenigl1991}.
The corotation radius is given by
\begin{equation}
r_{\rm co} = {\left( \frac{G M}{{\omega_{\rm spin}}^{2}} \right)}^{1/3} \simeq 5.3 R_{\rm J} {\left( \frac{2 \pi / \omega_{\rm spin}}{36\ {\rm hours}} \right)}^{2/3},
\end{equation}
where $\omega_{\rm spin}$ is the angular velocity of the planetary spin.
The rotation period of current Jupiter is 10 h; thus, the angular velocity of young Jupiter should have been about four times lower than that of current Jupiter if the inner cavity were created by the magnetic coupling between young Jupiter and the circumjovian disk.
In contrast, if $r_{\rm co} > r_{\rm in}$, a steady-state accretion disk cannot form, but a disk with no accretion might form.
Instead, the spin rate of the central planet would decrease by propeller-driven outflows \citep[e.g.,][]{Romanova+2005}.
Then, the spin-down of the planet increases the corotation radius, reaching a stable state with $r_{\rm co} \simeq r_{\rm in}$, as \citet{Sasaki+2010} predicted.

If we model the interior structure of the young Jupiter as a polytropic body with a polytropic index of 1.5, the spin angular momentum $L_{\rm spin}$ is given by \citep{Batygin2018}
\begin{equation}
L_{\rm spin} \simeq 0.21 M R^{2} \omega_{\rm spin}.
\end{equation}
Then, the planetary radius of young Jupiter would be as large as twice that of current Jupiter if we assumed conservation of spin angular momentum.
Evolutionary calculations of the planetary interior demonstrated, however, that the planetary radius of young Jupiter should have been about 1.5 times the current radius at 1 Myr after the formation of Jupiter \citep[e.g.,][]{Fortney+2011}.
Therefore, we need to consider the significant radius anomaly of young Jupiter in this scenario.

We note that \citet{Sasaki+2010} argued that Io could form at its current position, even if the inner cavity was at the current corotaion radius ($r = 2.25 R_{\rm J}$); however, subsequent $N$-body simulations by \citet{Ogihara+2012} revealed that the positions of the cavity and the innermost satellite must have been the same.

\section{Radiation pressure}
\label{AppC}

The radiation pressure is caused by the direct momentum transfer of photons.
The radiation force $F_{\rm rad}$ is given by
\begin{equation}
F_{\rm rad} = \frac{I}{c} \pi a^{2},
\end{equation}
where $c = 3.00 \times 10^{10}\ {\rm cm}\ {\rm s}^{-1}$ is the speed of light.
Here, we compare the photophoretic force $F_{\rm Ph}$ and the radiation force $F_{\rm rad}$.
The force ratio of the radiation force to the photophoretic force is
\begin{equation}
\begin{split}
\frac{F_{\rm rad}}{F_{\rm Ph}} &\simeq \frac{6 k_{\rm th} T}{c a p} \\
                               &\simeq 6 \times 10^{-2} {\left( \frac{T}{300\ {\rm K}} \right)} {\left( \frac{a}{3\ {\mu}{\rm m}} \right)}^{-1} {\left( \frac{p}{10^{2}\ {\rm dyn}\ {\rm cm}^{-2}} \right)}^{- 1}.
\end{split}
\end{equation}
Then, we found that the radiation force is negligibly weaker than the photophoretic force in the gaseous circumplanetary disks \citep[see also][]{Loesche+2012}.

\section{Mean free path of gas molecules}
\label{AppD}

\begin{figure}
\centering
\includegraphics[width = \columnwidth]{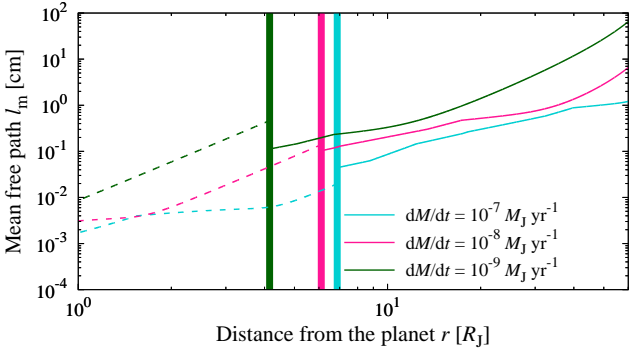}
\caption{
The mean free path of gas molecules in the disk.
The solid lines correspond to the mean free path $l_{\rm m}$ calculated from the normal model, and the dashed lines are the dust-poor model.
}
\label{figL}
\end{figure}

The mean free path of the gas molecules $l_{\rm m}$ is given by
\begin{equation}
l_{\rm m} = \frac{m_{\rm g}}{\sigma_{\rm mol} \rho_{\rm g}},
\end{equation}
where $\sigma_{\rm mol} = 2 \times 10^{-15}\ {\rm cm}^{2}$ is the collisional cross section of gas molecules \citep{Adachi+1976}.
In the circumjovian disk, $l_{\rm m}$ is larger than $10^{-2}\ {\rm cm}$ at the location of the light barrier (see Figure \ref{figL}), and the photophoretic force for small dust particles of $a = 3\ {\mu}{\rm m}$ is given by Equation (\ref{eq1}), as we assumed.

For spherical particles whose radius is smaller than the mean free path of the gas molecules, $a < l_{\rm m}$, the photophoretic force $F_{\rm Ph}$ is given by \citep{Beresnev+1993,Wurm+2006}
\begin{equation}
F_{\rm Ph} \simeq \frac{\pi a^{3} I p}{6 k_{\rm th} T},
\end{equation}
then $F_{\rm Ph}$ does not depend on $l_{\rm m}$.
On the other hand, when the dust radius is larger than the mean free path of the gas molecules, $a > l_{\rm m}$, the photophoretic force is approximately given by \citep{Takeuchi+2008}
\begin{equation}
F_{\rm Ph} \simeq \frac{\pi a^{3} I p}{6 k_{\rm th} T} \times 4.5 {\left( \frac{l_{\rm m}}{a} \right)}^{2},
\end{equation}
and $F_{\rm Ph}$ decreases with increasing of $a$.
In contrast to $F_{\rm Ph}$, the residual gravity caused by gas drag $F_{\rm D}$ is independent of $a / l_{\rm m}$ \citep{Weidenschilling1977}:
\begin{equation}
F_{\rm D} = - \frac{4 \pi a^{3}}{3} \frac{\rho_{\rm p}}{\rho_{\rm g}} \frac{{\rm d}p}{{\rm d}r}.
\end{equation}
Therefore, $F_{\rm Ph} / F_{\rm D}$ is independent of $a$ for the case of $a < l_{\rm m}$, while $F_{\rm Ph} / F_{\rm D}$ is inversely proportional to the square of the particle radius for the case of $a > l_{\rm m}$

This means that the radial drift of cm-sized compact pebbles could not be prevented by photophoresis itself, although the radial drift would be stopped because the pressure maximum formed at the light barrier.
We note, however, that the thermal conductivity of dust aggregates is exceedingly lower than that of compact dust particles \citep[e.g.,][]{Arakawa+2017,Arakawa+2019}, and the radial drift of aggregated pebbles might be stopped by photophoresis, even if their radius is larger than the mean free path of the gas molecules.

\section{Rotation of dust particles}
\label{AppE}

If particles rotate faster than the heat conduction timescale, the temperature gradient would be reduced, and the photophoretic force might be suppressed.
For $\mu$m-sized dust particles, the main source of particle rotation is the Brownian motion.
The thermal rotation timescale of a spherical particle, $\tau_{\rm rot}$, is \citep{Krauss+2005,Loesche+2012}
\begin{equation}
\tau_{\rm rot} = \sqrt{\frac{8 {\pi}^{2} a^{5} \rho_{\rm p}}{45 k_{\rm B} T}} \simeq 1.4 \times 10^{-2} {\left( \frac{a}{3\ {\mu}{\rm m}} \right)}^{5/2} {\left( \frac{T}{300\ {\rm K}} \right)}^{- 1/2}\ {\rm s}.
\end{equation}
The heat conduction timescale within a dust particle, $\tau_{\rm cond}$, is \citep{Krauss+2005,Loesche+2012}
\begin{equation}
\tau_{\rm cond} = \frac{a^{2}}{k_{\rm th}} \rho_{\rm p} c_{\rm p} \simeq 6.0 \times 10^{-5} {\left( \frac{a}{3\ {\mu}{\rm m}} \right)}^{2}\ {\rm s},
\end{equation}
where $c_{\rm p} = 10^{7}\ {\rm erg}\ {\rm g}^{-1}\ {\rm K}^{-1}$ is the specific heat \citep[e.g.,][]{Ciesla+2004a}.
We found that $\tau_{\rm rot} \gg \tau_{\rm cond}$, and the effect of particle rotation is negligible for $\mu$m-sized dust particles.

In addition to Brownian motion, dust-dust collisions spin up dust particles.
However, this effect is negligibly small for a dust-to-gas mass ratio lower than unity \citep[see][]{McNally+2017}.

\section{Dust and gas opacities}
\label{AppF}

We calculated the opacity of the disk using standard temperature-dependent opacity models \citep{Henning+1996,Freedman+2014}.

The contribution of dust to the Rosseland mean opacity of a disk with solar metallicity, $\kappa_{\rm R, d}$, is given by \citet{Henning+1996}, and we use the iron-poor model of the study.
Then, the contribution of dust to the Rosseland mean opacity of a disk with solar metallicity, $\kappa_{\rm P, d}$, is given by $\kappa_{\rm P, d} = 2.4 \kappa_{\rm R, d}$ \citep[][]{Nakamoto+1994}.

\begin{figure}
\centering
\includegraphics[width = \columnwidth]{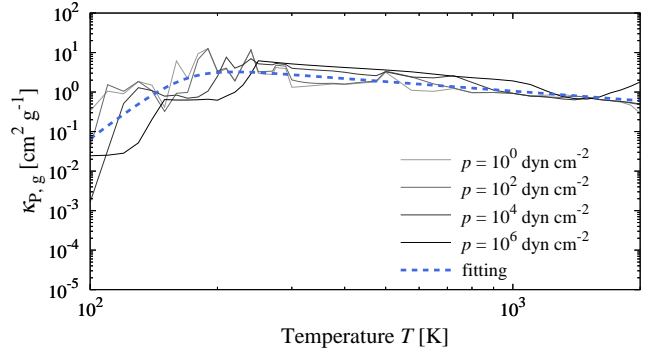}
\caption{
The Planck mean opacity of gas $\kappa_{\rm P, g}$ calculated by \citet{Freedman+2014} (gray solid lines) and the analytical fit used in this study (blue dashed line).
}
\label{fig_app}
\end{figure}

The contribution of gas to the Rosseland mean opacity of a disk with solar metallicity, $\kappa_{\rm R, g}$, is given by \citet{Freedman+2014}, and we use their analytical fit of $\kappa_{\rm R, g}$ for the solar metallicity gas.
We also evaluated $\kappa_{\rm P, g}$, using Table 3 of \citet{Freedman+2014} as
\begin{equation}
\kappa_{\rm P, g} = {\left[ 15 {\left( T / {100\ {\rm K}} \right)}^{-8} + 0.15 {\left( T / {100\ {\rm K}} \right)}^{0.8} \right]}^{-1}\ {\rm cm}^{2}\ {\rm g}^{-1}.
\end{equation}
Within the temperature range of 100--2000 K, the pressure dependence of $\kappa_{\rm P, g}$ is small.
Thus, we assume $\kappa_{\rm P, g}$ is only the function of the temperature (see Figure \ref{fig_app}).

We note that the dominant contributions to the Rosseland mean opacity of gas are in opacity windows, while the dominant contributions to the Planck mean opacity of gas are strong bands \citep[see Figure 2 of][]{Freedman+2014}.
Therefore, $\kappa_{\rm P, g}$ is several orders of magnitude larger than $\kappa_{\rm R, g}$ in general.

\section{Dependence on the turbulent viscosity}
\label{AppG}

In the main text, we assumed $\alpha = 10^{-3}$ for the outer region and $\alpha = 10^{-2}$ for the inner region.
However, the choice of absolute values of $\alpha$ does not change the results significantly, but the increase in $\alpha$ at the light barrier is the key to generate the bump in the surface density distribution.

\begin{figure}
\centering
\includegraphics[width = \columnwidth]{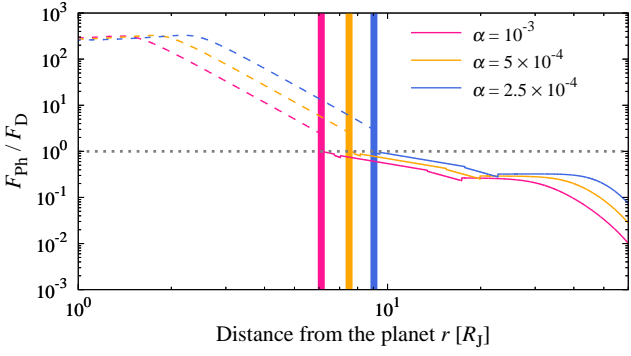}
\caption{
The force ratio $F_{\rm Ph} / F_{\rm D}$.
The solid lines correspond to the force ratio calculated from the normal model, and the dashed lines are the dust-poor model.
The vertical lines represent the location of the light barrier.
The line colors represent the value of $\alpha$ for the outer region ($\alpha = 10^{-3}$ for pink, $\alpha = 5 \times 10^{-4}$ for orange, and $\alpha = 2.5 \times 10^{-4}$ for blue).
We assume that the value of $\alpha$ for the inner region is 10 times higher than that for the outer region, and the mass accretion rate is $\dot{M} = 10^{-8} M_{\rm J}\ {\rm yr}^{-1}$.
}
\label{figFweak}
\end{figure}

Figure \ref{figFweak} shows the force ratio $F_{\rm Ph} / F_{\rm D}$ for various viscosity parameters.
Here, we assume that the value of $\alpha$ for the inner region is 10 times higher than that for the outer region, and the mass accretion rate is $\dot{M} = 10^{-8} M_{\rm J}\ {\rm yr}^{-1}$.
The line colors represent the value of $\alpha$ for the outer region ($\alpha = 10^{-3}$ for pink, $\alpha = 5 \times 10^{-4}$ for orange, and $\alpha = 2.5 \times 10^{-4}$ for blue).
We found that the location of the light barrier is $r_{\rm lb} = 6.1 R_{\rm J}$, $7.5 R_{\rm J}$, and $9.0 R_{\rm J}$ for the cases of $\alpha = 10^{-3}$, $5 \times 10^{-4}$, and $2.5 \times 10^{-4}$, respectively.
The location of the light barrier is not strongly dependent on $\alpha$ and we can adjust $r_{\rm lb} = 5.9 R_{\rm J}$ by choosing appropriate values of the luminosity $L$ and mass accretion rate $\dot{M}$.

\begin{figure}
\centering
\includegraphics[width = \columnwidth]{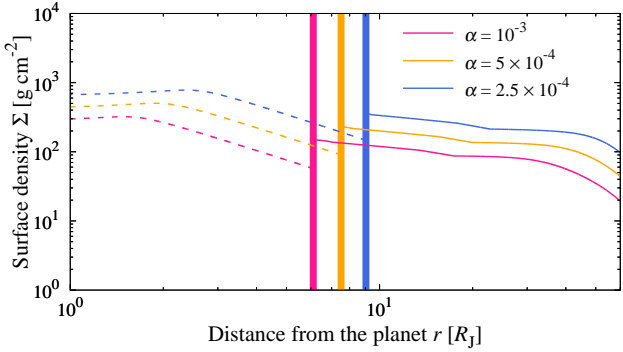}
\caption{
The surface density distribution of the disk for various viscosity parameters.
The solid lines correspond to the surface density $\Sigma$ calculated from the normal model, and the dashed lines are the dust-poor model.
The vertical lines represent the location of the light barrier, and a bump in the surface density distribution is formed at the light barrier.
}
\label{figSweak}
\end{figure}

Figure \ref{figSweak} shows the surface density distribution of the disk for various viscosity parameters.
We confirmed that a bump in the surface density distribution forms at the light barrier in all cases.
Therefore, the general view of the disk structure hardly depends on the choice of absolute values of $\alpha$, provided that the turbulence of the inner region is stronger than that of the outer region.

\section{Dependence on the particle radius}
\label{AppH}

For $\mu$m-sized dust particles in the circumjovian disk, we can simplify Equation (\ref{eq1}) as follows:
\begin{equation}
F_{\rm Ph} \simeq \frac{\pi a^{3} I p}{6 k_{\rm th} T}.
\end{equation}
Thus the force ratio $F_{\rm Ph} / F_{\rm D}$ is given by
\begin{equation}
\frac{F_{\rm Ph}}{F_{\rm D}} \simeq \frac{\rho_{\rm g} I p}{8 \rho_{\rm p} k_{\rm th} T} {\left( \frac{{\rm d}p}{{\rm d}r} \right)}^{-1},
\end{equation}
and the location of the light barrier $r_{\rm lb}$ is independent of the particle radius $a$, because both $F_{\rm Ph}$ and $F_{\rm D}$ are proportional to the cubic of $a$.

\section{Condition for resonance capture}
\label{AppI}

The Galilean satellites are believed to have migrated in the circumjovian disk by exchanging their angular momentum with the disk.
If the migration timescale, $\tau_{\rm mig}$, is longer than the critical migration timescale for resonance capture, $\tau_{\rm crit}$, the satellites are captured in the 2:1 mean-motion resonance.
\citet{Ogihara+2013} performed $N$-body simulations and revealed that the critical migration timescale for resonance capture $\tau_{\rm crit}$ is given by
\begin{equation}
\tau_{\rm crit} \simeq 2 \times 10^{5} {\left( \frac{M_{\rm inner}}{M_{\rm Io}} \right)}^{- 4/3} {\Omega_{\rm K}}^{-1},
\end{equation}
where $M_{\rm Io} = 8.9 \times 10^{25}\ {\rm g}$ is the mass of Io and $M_{\rm inner}$ is the mass of the inner satellite.
The migration timescale $\tau_{\rm mig}$ is given by \citep[e.g.,][]{Paardekooper+2011,Fujii+2017}
\begin{equation}
\begin{split}
\tau_{\rm mig} &\simeq \frac{1}{5} {\left( \frac{M_{\rm sat}}{M} \right)}^{-1} {\left( \frac{r^{2} \Sigma}{M} \right)}^{-1} {\left( \frac{c_{\rm s}}{r \Omega_{\rm K}} \right)}^{2} {\Omega_{\rm K}}^{-1} \\
               &\simeq 7 \times 10^{7} {\left( \frac{T}{300\ {\rm K}} \right)} {\left( \frac{\Sigma}{300\ {\rm g}\ {\rm cm}^{-2}} \right)}^{-1} {\left( \frac{r}{9 R_{\rm J}} \right)}^{-1} {\left( \frac{M_{\rm sat}}{M_{\rm Eu}} \right)}^{-1} {\Omega_{\rm K}}^{-1},
\end{split}
\end{equation}
where $M_{\rm Eu} = 4.8 \times 10^{25}\ {\rm g}$ is the mass of Europa.
We found that $\tau_{\rm mig} \gg \tau_{\rm crit}$ and, therefore, the adjacent satellites (i.e., Io--Europa and Europa--Ganymede) can be captured in the 2:1 mean motion resonance.

The migration timescale $\tau_{\rm mig}$ is on the order of a few $10^{5}$ years for the Galilean satellites, and they can migrate in the circumjovian disk.
We note, however, that $\tau_{\rm mig}$ might be longer than the dissipation timescale of the disk when the mass of the satellite is far smaller than $10^{-2} M_{\rm Eu}$.
This fact indicates that Europa and Ganymede should be larger than $10^{-2}$ times the current mass at the time of resonance capture.

\end{appendix}

\end{document}